\documentclass[12pt,twoside]{article}   
\usepackage[super,sort,comma]{natbib}

\usepackage{amsmath}

\usepackage{amsfonts}
\usepackage{amssymb}
\usepackage{float}
\usepackage{listings}
\usepackage{gensymb}
\usepackage{eurosym}
\usepackage{caption}
\usepackage{subcaption}
\usepackage{xfrac}
\usepackage{array}

\usepackage{fancyhdr}		

\usepackage[section]{placeins}   %

\usepackage{graphicx}

\makeatletter \renewcommand\@biblabel[1]{$^{#1}$} \makeatother
\newcommand{\cen}[1]{\begin{center} #1 \end{center}}

\usepackage[mathlines]{lineno}

\usepackage{hyperref}
\hypersetup{ colorlinks,
    citecolor=blue,
    filecolor=blue,
    linkcolor=blue,
    urlcolor=blue
}

\usepackage{xcolor}

\definecolor{gray}{rgb}{0.6,0.6,0.6}
\definecolor{red}{rgb}{0.85,0,0}
\definecolor{green}{rgb}{0,0.85,0}
\definecolor{blue}{rgb}{0,0,0.85}
\definecolor{beige}{rgb}{0.92,0.87,0.78}
\usepackage[all]{hypcap}    

\begin{document}

\cen{\sf {\Large {\bfseries Real-time image-guided treatment of mobile tumors in proton therapy by a library of treatment plans: a simulation study} \\  
\vspace*{10mm}
Valentin Hamaide\textsuperscript{a}, Kevin Souris\textsuperscript{b}, Damien Dasnoy\textsuperscript{a}, François Glineur\textsuperscript{a}, Benoît Macq\textsuperscript{a}} \\
\textsuperscript{a}ICTEAM Institute, UCLouvain, Louvain-la-Neuve, Belgium\\
\textsuperscript{b}MIRO Institute, UCLouvain, Brussels, Belgium
\vspace{5mm}\\
\today\\
}

\pagenumbering{roman}
\setcounter{page}{1}
\pagestyle{plain}
Corresponding author: Valentin Hamaide. email: valentin.hamaide@uclouvain.be

\begin{abstract}
\noindent  {\bf Purpose:} To improve target coverage and reduce the dose in the surrounding organs-at-risks (OARs), we developed an image-guided treatment method based on a precomputed library of treatment plans controlled and delivered in real-time.\\
{\bf Methods:} A library of treatment plans is constructed by optimizing a plan for each breathing phase of a 4DCT. Treatments are delivered by simulation on a continuous sequence of synthetic CTs generated from real MRI sequences. During treatment, the plans for which the tumor is at a close distance to the current tumor position are selected to deliver their spots. The study is conducted on five liver cases.\\
{\bf Results:} We tested our approach under imperfect knowledge of the tumor positions with a 2 mm distance error. On average, compared to a 4D robustly optimized treatment plan, our approach led to a dose homogeneity increase of 5\% (defined as $1-\frac{D_5-D_{95}}{\text{prescription}}$) in the target and a mean liver dose decrease of 23\%. The treatment time was roughly increased by a factor of 2 but remained below 4 minutes on average. \\
{\bf Conclusions:} Our image-guided treatment framework outperforms state-of-the-art 4D-robust plans for all patients in this study on both target coverage and OARs sparing, with an acceptable increase in treatment time under the current accuracy of tumor tracking technology.

\end{abstract}

\newpage     
\tableofcontents

\newpage

\setlength{\baselineskip}{0.7cm}      

\pagenumbering{arabic}
\setcounter{page}{1}
\pagestyle{fancy}
\section{Introduction}

\noindent Proton therapy offers a physical advantage over conventional radiotherapy in terms of dose conformity and normal tissue sparing due to the unique depth-dose characteristics of protons, thus potentially improving tumor control while at the same time reducing toxicity \cite{mohan2017proton}. However, this precision comes at the cost of high vulnerability to uncertainties. This is especially the case for thoracic tumors, where breathing-induced tumor motion can lead to density variation in the beam path, thereby missing the target or shifting the expected proton range and deteriorating the overall dose distribution. Moreover, with pencil beam scanning (PBS) proton therapy, on top of the proton range variation issue, another dose deterioration can occur due to the interference between the scanning beam motion and the anatomical motion, known as the interplay effect, which can lead to hot and cold spots in the target \cite{seco2009breathing}.\\
\\
The current state-of-the-art treatment planning approach for dealing with intra-fractional motion is 4D robust optimization which is designed to be robust against small changes in the anatomy by optimizing the worst-case scenario. On top of the usual 3D robustness scenarios, such as range and setup errors, 4D robust optimization is designed to be robust against multiple anatomical variations present in the 4DCT. While this approach is efficient and is the current best practice for treating thoracic cancer \cite{chang2017consensus}, two criticisms can be made: (i) the treatment is only designed to be robust against motions seen in the planning 4DCT, and (ii) robustness against breathing phases increase the dose to the surrounding organs at risks (OARs).\\
\\
Another passive motion mitigation method is rescanning, which refers to delivering the dose in multiple iterations to smooth out the positional errors due to the interplay effect \cite{phillips1992effects}. While rescanning is interesting to mitigate the interplay, it does not solve the range variation issue. When the tumor motion amplitude is high, active techniques such as gating, abdominal compression, or breath-hold using optical or fluoroscopy imaging are generally used \cite{czerska2021clinical}. A breath-hold approach has the advantage of being simple and easily implemented, with reduced tumor motion and increased tumor-to-OAR distance \cite{hanley1999deep}. However, this method requires coaching, and not all patients can hold their breath for a long enough period. Abdominal compression is also a relatively simple method to implement and has been shown to reduce tumor motion \cite{souris2016f}. However, it only mitigates the dose deterioration to a certain degree. Another common active motion mitigation technique is respiratory gating which switches on the beam only during a specific respiratory window based on the phase or amplitude of a breathing signal \cite{lu2007respiratory}. This has the advantage of increasing dose conformation in the target and sparing OARs. However, this method requires synchronization with tumor motion, which is challenging and increases the treatment time. A motion mitigation approach not mentioned is beam tracking, which would be the optimal technique should it be implemented one day since it would not lead to target expansion and increased treatment time \cite{grozinger2006simulations}. However, beam tracking in proton therapy is very sensitive to position uncertainties and requires hardware adaptations to be able to switch the energy in a fast way to compensate for range changes \cite{van2009tumour}. Hence, tracking in particle therapy remains a research topic and will take some time before being implemented clinically \cite{mori2018motion}.\\
\\
Our proposed approach is somewhat an extension of the respiratory gating approach where the beam is ON only when the tumor position is at a close distance to one particular phase of the planning 4DCT. However, instead of optimizing one global treatment plan, our approach consists of optimizing ten separate treatment plans corresponding to each phase of the 4DCT. Then, during treatment, the plans associated with the planning CT phases whose tumor position is closest to the current tumor position deliver their spots.\\
\\
A similar idea was considered by Graeff \cite{graeff2014motion} and subsequently implemented in a prototype system at GSI Helmholtzzentrum für Schwerionenforschung \cite{lis2020modular}, where they also use a library of treatment plans. However, their delivery process is different. They take their decision based on a 1D signal from an external surrogate with a phase-based approach and irradiate in a continuous fashion. Our approach consists of taking decisions based on the frequent acquisition of images and the use of a particular distance metric between the current image and the 4DCT and irradiating in an intermittent fashion. A drawback of the phase-based approach is that the beam might miss the target because the target position in the phase space might not be the same, even though the corresponding respiratory phase is the same. In our method, we base our decision on the distance between the current tumor position and the tumor position in the planning 4DCT. The distance metrics tested in this paper are the Euclidean distance between the center of mass of the tumors and the DICE similarity index between the tumors. A maximum distance (or minimum DICE) threshold needs to be set by the user, which will trade off between dose conformity and treatment time. Several thresholds are compared in this paper and confronted with inexact tumor positions obtained by adding noise of various amplitudes.\\
\\
We compare state-of-the-art 4D robust treatment plans against our approach in Section \ref{sec:results} by simulating the treatment on a continuous sequence of synthetic 3DCTs generated from cine-MRI sequences of five liver patients. This has the benefit of being as close as possible to a real breathing scenario that can be encountered in practice and that does not necessarily follow the breathing pattern seen in the original 4DCT.

\section{Materials \& methods}
\subsection{Patient data}
\label{sec:patient_data}
\noindent The study includes five patients with liver cancer that were treated with photon radiotherapy for which a planning 4DCT was acquired with abdominal compression and audio-coaching to regularize the breathing pattern. Those patients were replanned in this study for proton therapy with the characteristics described in Section \ref{sec:planning}. The particularity of this study is that on top of the planning 4DCT, the patients underwent a dynamic MRI scan. The MRI sequence was acquired the same day and under the same conditions as the 4DCT, with abdominal compression and audio-coaching at a rate of 1.85Hz during 2 minutes. The 2D dynamic MRI sequence can then be used to generate a continuous sequence (CS) of synthetic 3DCTs with the method of Dasnoy et al. \cite{dasnoy2020continuous, dasnoy2022locally}. This continuous sequence represents a real breathing pattern of the patient, on which we can simulate the treatment and compute the accumulated dose. Patients' tumor information are detailed in Table \ref{tab:patient}. We can make several observations from those data. Patient 1's motion amplitude observed in the continuous sequence is much lower than the motion amplitude observed in the 4DCT. Indeed, the motion amplitude in the 4DCT overestimates the motion by roughly a factor of 3. Motion amplitudes for the rest of the patients remain globally similar on average (within 2-2.5mm) between what is observed in the 4DCT and the continuous sequence. However, the maximum peak-to-peak amplitudes can be quite different (e.g. for patients 4 and 5). Those differences in motion amplitudes are not uncommon. Indeed, in a recent study, the authors observed a maximum amplitude deviation from the 4DCT $>$ 5 mm for liver tumors in 67\% of the cases and a maximum amplitude deviation from the 4DCT $>$ 10 mm in 22\% of the cases in the CC direction \cite{dhont2018long}. However, the average amplitude deviation from the 4DCT $>$ 5 mm (and $>$ 10 mm) was observed for only 6\% of the liver tumors, so patient 1 shows a less common amplitude variation. \\

\begin{table}
    \small
    \resizebox{\textwidth}{!}{%
    \begin{tabular}{p{1.75cm}|p{1.35cm}|p{1.1cm}|p{1.1cm}|p{1.1cm}|p{2.55cm}|p{2.55cm}|p{2.55cm}}
         & Tumor volume [$cm^3$] & \multicolumn{3}{p{3.3cm}}{P2P motion amplitude in 4DCT [mm]} & \multicolumn{3}{|p{7.65cm}}{P2P motion amplitude in CS [mm]: Mean $\pm$ std (max)} \\
         & & LR & AP & CC & LR & AP & CC \\\hline
        Patient 1 & 29.73 & 7.5& 8.4& 10.7& 2.8$\pm$0.6 (4.2)& 2.7$\pm$0.6 (4.0)& 3.5$\pm$0.9 (5.0)\\
        Patient 2 & 9.41 & 1.2& 3.4& 7.9& 1.6$\pm$0.1 (2.0) & 3.1$\pm$0.2 (3.6)& 10.6$\pm$0.8 (12.7) \\
        Patient 3 & 8.12 & 0.8& 5.0& 8.6& 0.7$\pm$0.1 (0.8)& 4.3$\pm$1.5 (5.7)& 10.5$\pm$2.2 (13.1)\\
        Patient 4 & 82.40 & 1.1& 7.4& 7.8& 1.2$\pm$0.8 (3.8)& 3.5$\pm$2.9 (12.0)& 8.5$\pm$5.1 (19.8)\\
        Patient 5 & 22.62 & 2.6& 5.0& 11.3& 2.9$\pm$0.9 (5.8)& 5.4$\pm$1.1 (9.0)& 13.0$\pm$2.5 (20.3)
    \end{tabular}}
    \caption{Information on the tumor volume and peak-to-peak (P2P) motion amplitudes for the five patients in the case study. The motion amplitudes are given for both the 4DCT and continuous sequence (CS) in the three main directions: Left-right (LR) direction, Anterior-Posterior (AP) direction, and Cranio-Caudal (CC) direction.}
    \label{tab:patient}
\end{table}

\begin{figure}
\centering
\begin{subfigure}{\textwidth}
  \centering
  \includegraphics[width=\linewidth]{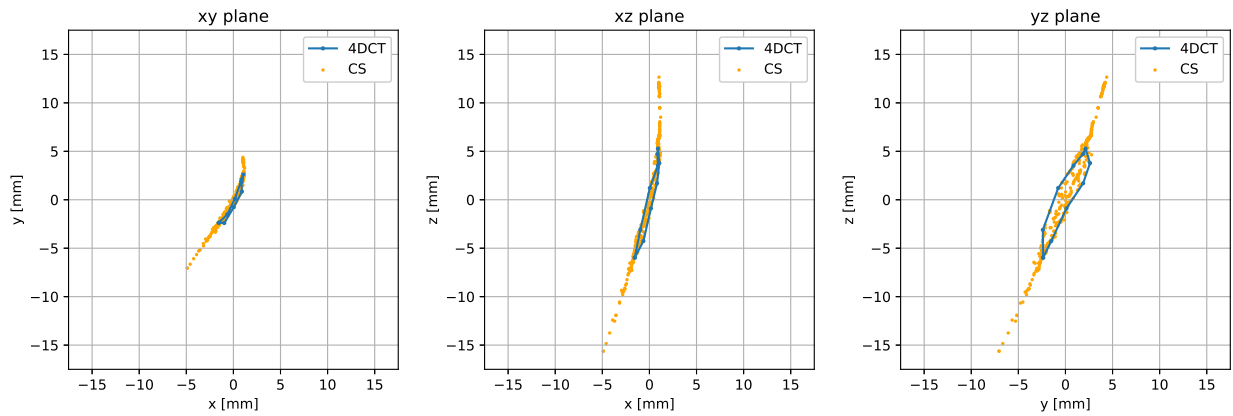}
  \caption{Motion of the tumor in two dimensions in the three principal planes (xy plane, xz plane, and yz plane).}
  \label{fig:motion_2d_p19}
\end{subfigure}
\begin{subfigure}{\textwidth}
  \centering
  \includegraphics[width=\linewidth]{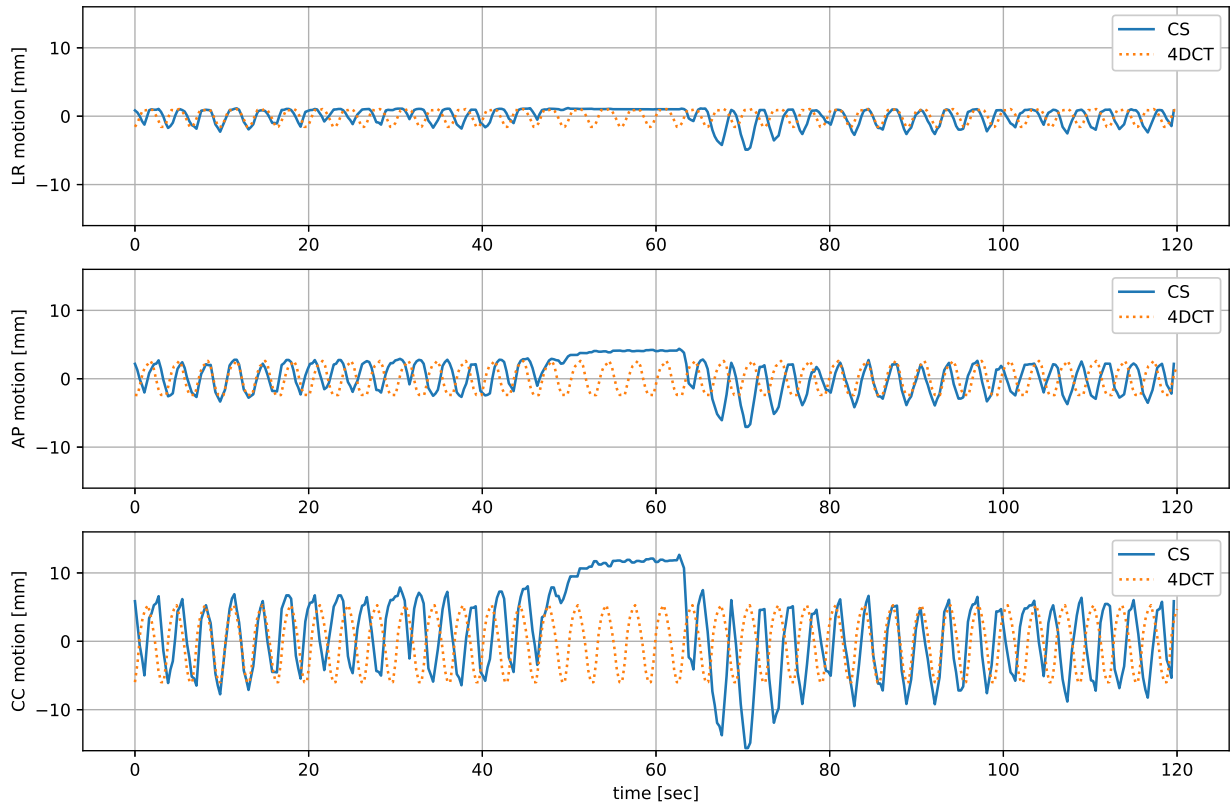}
  \caption{Motion of the tumor in 1 dimension as a function of time in the three principal directions: left-right (LF) direction, anterior-posterior (AP) direction, and cranio-caudal (CC) direction.}
  \label{fig:motion_1d_p19}
\end{subfigure}
\caption{Motion of the center of mass of the tumor in the 4DCT (looped over two minutes) and the continuous sequence (CS) for patient 5.}
\label{fig:motion_p19}
\end{figure}

\noindent An example of tumor motion that we can observe in one of the patients is depicted in Figure \ref{fig:motion_p19}. We observe that the tumor's location in the continuous sequence does not necessarily stay within the path formed by the 4DCT motion. Hence, some motions will not necessarily be taken into account even with a 4D robust optimized treatment plan. The tumor motions for the four other patients are available as supplementary materials for completeness (Figures S1-4).

\subsection{Treatment planning}
\label{sec:planning}
\noindent The conventional approach for designing a treatment plan is to encode desired physical constraints in an objective function to be minimized using a dose-influence matrix calculated on a reference CT image. Since proton dose deposition is subject to uncertainties due to density variations in the beam path and positional errors, we must make the optimization robust to those uncertainties by optimizing the worst-case scenario. Taking into account range errors and setup errors is generally referred to as 3D robust optimization, while additionally taking into account anatomical variations from the 4DCT is referred to as 4D robust optimization.\\
\\
In our approach, instead of optimizing a single treatment plan on a reference CT, we optimize 10 treatment plans, one on each phase of the 4DCT. There are several ways one can achieve that. The optimal way would be to concurrently optimize the ten plans so that the most optimal beam configuration is delivered in each phase to take advantage of their different tumor-to-OARs configurations. This approach leads to heterogeneous dose distribution in each plan that sums up to the theoretically best achievable dose distribution. Although this approach sounds appealing, it is a relatively dangerous idea as it would lead to a heterogeneous dose distribution if some anatomical configurations do not eventually appear during treatment, not even mentioning the additional computational burden of optimizing a huge plan while handling deformable image registration during the optimization. Instead, we propose to seek a homogeneous dose in each plan by optimizing a plan independently on each phase of the 4DCT with the same objective function, which is also one of the proposed methods by Graeff \cite{graeff2014motion}. The downside of this approach is that the total number of spots to deliver is increased since we optimize 10 plans instead of one, but this also leads to an intrinsic rescanning, which can mitigate the interplay effect.\\
\\
The library of treatment plans can be defined mathematically for each $\mathrm{CT}_i$ ($i=1,..,P$) in the planning 4DCT as
\begin{align}
    \min_{\mathbf{w}^{\mathrm{CT}_i}} \max_{s \in \mathcal{S}} &f(\mathbf{w}^{\mathrm{CT}_i}; \mathbf{B}^{\mathrm{CT}_i}_s)\label{eq:obj} \\
    w_j^{\mathrm{CT}_i} \geq P\cdot n_f\cdot m &\text{  OR  } w_j^{\mathrm{CT}_i}=0 \text{ for all $j\in \mathcal{J}$} \label{eq:constraint}
\end{align}
where $f$ is the objective function common to the ten problems, $\mathbf{B}^{\mathrm{CT}_i}_s$ is the beamlet matrix associated to the $i^{th}$ CT under scenario $s$ and $w_j^{\mathrm{CT}_i}$ the variables to optimize for that CT, which are the weights associated to beamlet indices $j\in \mathcal{J}$. This is the common worst-case robust optimization formulation for radiation therapy; only this has to be solved on each CT. This paper uses a set of scenarios $\mathcal{S}$ of $3\times 7=21$ scenarios: those combine 3 density variations (nominal and $\pm 3\%$) and 7 setup variations (nominal and a shift of $\pm2.5mm$ in each of the three dimensions). The weights associated to each beamlet must also either be zero or satisfy a lower bound constraint, which corresponds to the number of plans ($P=10$ in our case) multiplied by the number of fractions $n_f$ and the minimum number of monitor units $m$, the machine can deliver. Note that this lower bound constraint is more restrictive than for a conventional treatment plan (10 times greater) since each plan is designed to deliver 10\% of the dose in our case. However, this more restrictive constraint will lead to a sparser treatment plan since more weights will need to be set to zero by the optimizer to achieve higher weights elsewhere. The disjunctive constraint \eqref{eq:constraint} is handled in two steps. First, the optimization problem is solved for $\mathbf{w}\geq 0$, then only the spots satisfying $\mathbf{w}\geq P\cdot n_f\cdot m$ are kept, and the optimization is refined by applying $\mathbf{w}\geq P\cdot n_f\cdot m$ as a hard constraint to the spots kept from the first step.\\
\\
The physical constraints expressed as penalties in the objective function $f$ to guide the optimization are defined in this paper as follows:
\begin{itemize}
    \item Target: $\min \mathbf{d}_T\geq p$ and $\max \mathbf{d}_T\leq p$ where $\mathbf{d}_T$ is the dose on each voxel in the target and $p$ is the prescription, which in our paper is chosen to be $p=60$Gy.
    \item OARs: $\max \mathbf{d}_{OAR}\leq \frac{2}{3}p$ for all organs at risk.
\end{itemize}
Furthermore, we emphasize the target constraints by weighting the target constraints 10 times more than the OARs constraints. The constraints in the objective functions are usually encoded as asymmetric quadratic penalties, although this depends on the treatment planning system (TPS) used.\\
\\
Concerning the grid spacing parameter to solve equation \eqref{eq:obj}, we choose a constant spot spacing of 5 mm and a constant layer spacing of 5 mm (water equivalent distance). A special consideration for our approach is that the treatment plans need to share the same energy layers to avoid constantly switching energy during delivery, which would significantly increase the treatment time. Hence, after the first optimization is carried out, we enforce the next plans to reuse the same energy layers. Nevertheless, energy layers that would be needed outside the current range of energies in other plans can still be added, and energy layers not required in a plan can be removed in that plan. In the end, most energy layers will be shared across plans, with only a few used only by extreme phases.\\
\\
In Section \ref{sec:results}, we compare the static and dynamic dose delivered by a 4D robust plan against our approach. For the library of plans approach, the static dose is defined as the sum of optimized doses from each phase plan deformed on the reference CT. Mathematically, this is
\begin{align}
\label{eq:accumulation1}
    \mathbf{F}_i &= \mathrm{DIR}(\mathbf{CT}_i,\mathbf{CT}_{\text{MidP}}) \text{ for all $i=1,...,P$} \\
    \mathbf{D}_{\text{MidP}} &= \sum_{i=1}^{P}\mathbf{D}_i(\mathbf{x} + \mathbf{F}_i(\mathbf{x})) \text{ for all positions $\mathbf{x}\in\mathbb{R}^3$ in the image.}
    \label{eq:accumulation2}
\end{align}
where $\mathrm{DIR}(\cdot)$ is the deformable image registration operator, $\mathbf{F}_i$ is the deformation field resulting from the registration, $\mathbf{CT}_i$ is planning CT image $i$, $\mathbf{CT}_{\text{MidP}}$ is the Mid-position CT and $\mathbf{D}_i$ is the optimized dose computed on $\mathbf{CT}_i$. The reference CT is, in our case, the Mid-position (MidP) CT \cite{wolthaus2008reconstruction}. This gives the theoretically best achievable dose by our library of treatment plans.

\subsection{Treatment delivery}
\label{sec:delivery}
\begin{figure}
    \centering
    \includegraphics[scale=0.6]{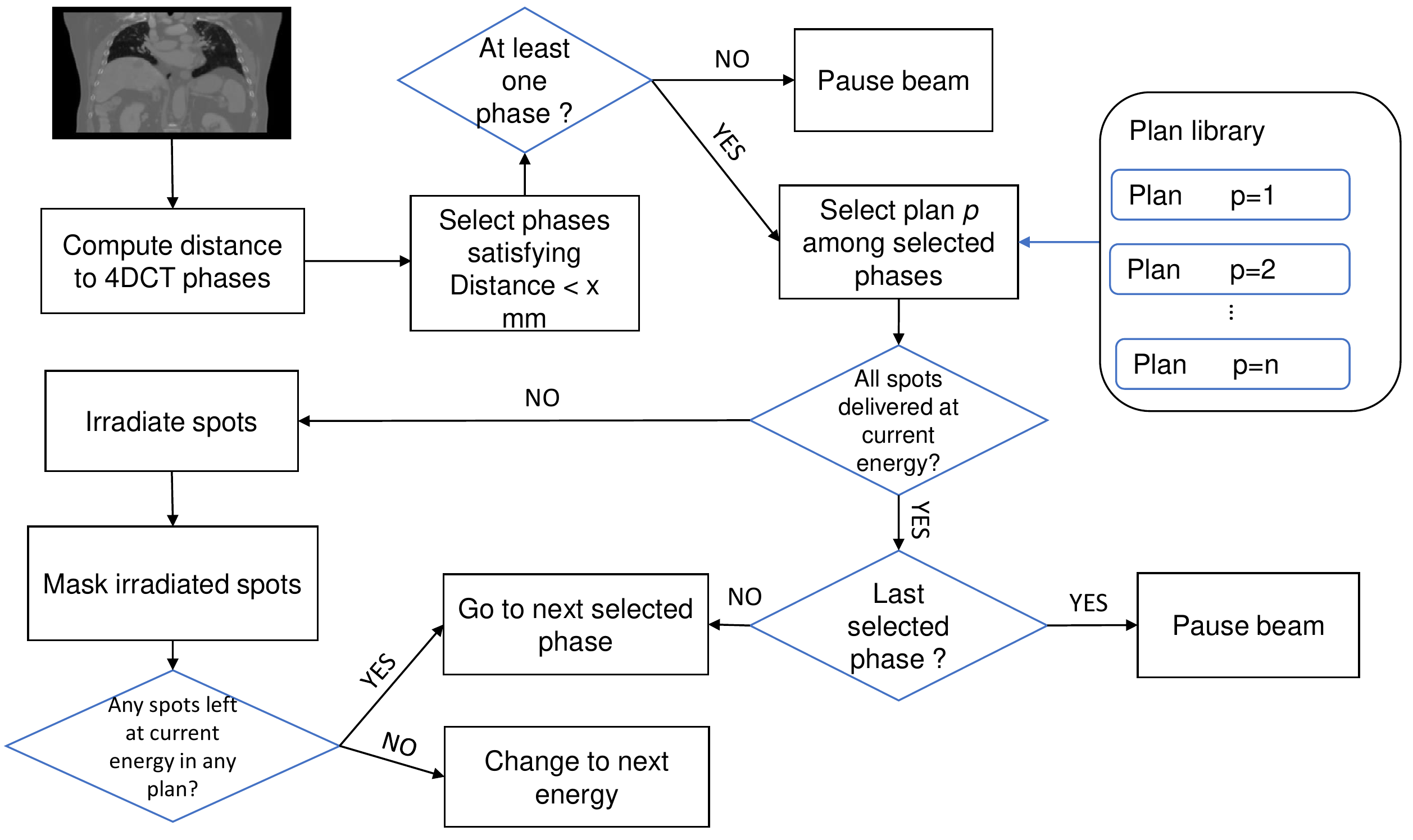}
    \caption{Treatment delivery process of our library of treatment plans approach.}
    \label{fig:delivery}
\end{figure}

\noindent Proton treatment plans are usually delivered without any synchronization, i.e. all the spots are delivered in a continuous fashion layer by layer in a serpentine pattern regardless of the changes in anatomy until all spots are delivered. Our approach is an active technique that irradiates only when the current target position is close to its position in a phase of the planning 4DCT. It is essentially an improved version of respiratory gating with multiple plans computed on each 3DCT. A flowchart representing the treatment delivery framework is depicted in Figure \ref{fig:delivery}.\\

\noindent First, an image is acquired from which the tumor's location is retrieved. Our approach is agnostic of the modality used to obtain the tumor location information. This could be via fluoroscopy imaging with a fiducial marker \cite{miyamoto2019quantitative}, markerless tracking \cite{hirai2019real}, or even magnetic resonance imaging. As long as we receive accurate information on the location of the tumors a few times every second, our approach is applicable. We could also eventually use an external surrogate with a tumor correlation model, although the indirect measurements make this solution less accurate. Hence, we suppose that we get a possibly noisy observation of the tumor's location. We then compute a distance between the current tumor position and the tumor positions in each phase of the 4DCT. Two distance metrics are detailed in Section \ref{sec:distance_measure_method} for this purpose. The phases in which the target position is closer to the current target position than a certain threshold are selected, and the plans deliver their spots at the current energy until the next image is acquired or until no more spots remain in the selected plans at that particular energy. The order in which those plans are delivered depends on their completion progress, i.e. the plan with the most remaining spots will be delivered first. The reason for prioritizing the less delivered plan instead of the more similar one is to decrease treatment time. Once a new image is acquired, the process restarts. When all plans have delivered all their spots at the current energy, the energy is lowered to the next energy layer. If no plan satisfies the distance constraint, the beam is paused until the next image acquisition. Finally, to avoid unreasonable treatment time, if no spots were irradiated for more than 10 consecutive seconds, the distance threshold is gradually increased. In our simulation, we choose to increase the distance threshold to 5mm after 10 seconds and 10 mm after 20 seconds and decrease it again to the original distance threshold after the current energy layer is completed.

\paragraph{Initialization phase}\; \\
The tumor's motion observed in the planning 4DCT might be different than the one observed during the day of the treatment. There might be a difference in breathing amplitude, or there could be a baseline shift \cite{zeng2022analysis}. To account for the true breathing pattern of the day, we implemented an initialization phase during which we observe the tumor motion 30 seconds before starting the beam. During this period, images are acquired in the same way they are acquired during treatment, i.e. a few images per second. For each image, the tumor's location is computed, and the distances to the ones from the 4DCT are also computed. The phases whose distance is below the user-defined threshold are recorded for all images in the initialization phase. Finally, only the phases that appear regularly are kept for the treatment delivery, and their weights are rescaled accordingly to meet the dose prescription. The occurrence of phase $p_i$ during the initialization phase is computed as follows:
\begin{equation}
\label{eq:Opi}
    O_{p_i} = \frac{100}{N} \cdot \# \left \lbrace \mathrm{dist}\left(c\left(\mathrm{im}_j\right), c(p_i)\right) < D, i=1,...,N \right \rbrace
\end{equation}
where $N$ is the number of images acquired during the initialization period, $c(\cdot)$ is a function computing the location of the center of mass of the tumor, $\mathrm{dist}$ is the Euclidean distance, $\mathrm{im}_j$ is the image $j$ acquired during the initialization phase and $\#\lbrace \cdot \rbrace$ is the counting operator. If $O_{p_i} > 5\%$, the plan corresponding to phase $p_i$ is selected for the treatment delivery, that is with at least half of the \emph{normal} occurrence of a phase in the 4DCT.\\
\\
After the selection process, the weights of the selected plans must be rescaled to reach the prescription dose. Let $n_p$ be the number of selected plans from the initialization phase. Then the weights of each of the selected plans are rescaled as follows:
\begin{equation}
    \mathbf{w}^{CT_i} \leftarrow \frac{10}{n_p}\mathbf{w}^{CT_i}
\end{equation}
Indeed, since the dose in each plan is homogeneous, we can simply rescale all the weights by a constant so that their sum reaches the prescription dose.

\subsection{Choosing the distance metric}
\label{sec:distance_measure_method}
\noindent What information do we need or should we use for deciding which plan to choose for irradiating the patient at time $t$? A straightforward measure to use is the Euclidean distance between the center of mass of the tumor in the current image and the 4DCT. In this study, we consider the distance between the center of mass of the tumors, but this could also very well be the location of a fiducial marker for marker-based tracking. The most accurate information is naturally 3-dimensional, but we can also consider a 2-dimensional or even 1-dimensional location which would include the principal amplitude variations. In this study, we consider a 3D information with various noise amplitudes. Another theoretically more accurate distance metric is the DICE coefficient. The DICE similarity coefficient, first introduced by Dice \cite{dice1945measures} and subsequently adapted for segmentation \cite{zijdenbos1994morphometric}, is used as a measure of similarity between two binary masks, i.e. the segmentation of the tumor in the current image and the segmentation of the tumor in each phase of the 4DCT. It gives a score between zero and one (0 meaning no overlap and 1 means complete overlap) and is computed as 
\begin{equation}
    \text{DICE} = \frac{2|A \cap B|}{|A|+|B|}
\end{equation}
where $A$ and $B$ are two binary masks. The two metrics are compared in Section \ref{sec:distance_measure}.

\subsection{What image acquisition frequency do we need?}
Every time we receive an image, we can make a decision: whether or not to irradiate and which plans to shoot. Therefore, we need to know the optimal frequency to acquire images. Too few acquisitions would lead to less accurate and potentially longer treatments, while too many acquisitions might be undesirable in terms of radiation exposure for imaging modalities such as fluoroscopy and even unnecessary in terms of performance. Indeed, the tumor movement becomes negligible at smaller time scales. On top of the decision-making, the frequency rate has an impact on the accuracy of the treatment delivery simulation. Different acquisition frequencies are tested in Section \ref{sec:image_frequency}, some requiring upsampling the continuous sequence of CTs. The upsampling is done by applying a deformation field, computed from each pair of images that follow each other in time. Mathematically, that is
\begin{align}
    \mathbf{F}_i &= \mathrm{DIR}(\mathbf{CT}_i,\mathbf{CT}_{i+1}) \\
    \mathbf{V}_i &= \log(\mathbf{F}_i + Id)\\
    \mathbf{CT}_{i+c} &= \mathbf{CT}_i(\mathbf{x} + \exp(c\mathbf{V}_i(\mathbf{x})) \text{ for all $\mathbf{x}\in\mathbb{R}^3$}
\end{align}
where $\mathbf{CT}_i$ is image $i$ and $\mathbf{CT}_{i+c}$ is an interpolated image between $i$ and $i+1$ with $c \in ]0,1[$ a constant, $\mathbf{F}_i$ and $\mathbf{V}_i$ the displacement and velocity field between phase $i$ and $i+1$ respectively, and $Id$ the identity transform ($Id(\mathbf{x})=\mathbf{x}$). The log and exp are operations on the vector fields (rather than element-wise operations) as described in Vercauteren et al. \cite{vercauteren2009diffeomorphic}.

\subsection{Simulation}
Treatment plans were designed with Raystation 10B \cite{bodensteiner2018raystation}, while the simulation of the delivery of the treatment plans on the continuous sequence of synthetic CT was carried out in our in-house treatment planning system. The computation of the spot delivery timings was done with the IBA ScanAlgo simulation tool emulating the delivery timings on an IBA C230 cyclotron, while the simulation of the dose deposition on the continuous sequence of 3DCTs was done with the Monte Carlo dose engine MCsquare \cite{souris2016fast}, which was recently validated for clinical use \cite{deng2020integrating}. The dose distributions in the CTs of the continuous sequence are then accumulated on the MidP CT via deformable image registration, similarly to equations (\ref{eq:accumulation1}-\ref{eq:accumulation2}).\\
\\
We compare the dose deposited by a 4D-robust plan against our library of treatment plans method, both simulated on a continuous sequence of around 400 3DCTs (see Section \ref{sec:patient_data}) representing 2 minutes of breathing. If the duration of the treatment is longer than 2 minutes, we assume that the breathing restarts in a loop from the first image. Our method is tested under various distance thresholds and noise levels as well as different starting points on the continuous sequence to assess the statistical variability. We assess the performance of our method based on three complementary metrics: the homogeneity in the target, the mean liver-CTV dose, and the treatment time. Homogeneity is defined as 
\begin{equation}
\label{eq:homogeneity}
    \text{homogeneity} = 1 - \frac{D_5-D_{95}}{\text{prescription}}
\end{equation}
where $D_{95}$ and $D_5$ are the lowest dose received at least 95\% and 5\% of the target volume respectively, and the prescription is in this study fixed at 60Gy. This metric gives a score between 0 and 1, where 1 represents a (nearly) perfect homogeneity in the target. The OAR sparing is in this study expressed as the mean liver-CTV dose where liver-CTV represents the volume subtraction between the liver and the CTV. Indeed, according to the consensus report from the Miami Liver Proton Conference, minimizing the mean liver dose and the volume of uninvolved tumor is of extreme importance for any liver radiotherapy \cite{chuong2019consensus}. Finally, the treatment time must also be taken into account as a trade-off with the target coverage.

\paragraph{Simulating noise on the location of the tumor}
To simulate imperfect information on the position of the tumor, we add Gaussian noise to the location of the center of mass of the tumor taken from the images of the continuous sequence. The noise is designed to represent an average distance error to the actual true position. Those noises are designed to follow a multivariate normal distribution $\mathbf{e}\sim \mathcal{N}(\mathbf{0},\boldsymbol{\Sigma})$ of zero mean and covariance matrix $\boldsymbol{\Sigma}=\sigma^2\mathbf{I}$ with $\sigma=Cd$ where $d$ represents the average distance error and $C=\frac{\sqrt{\pi}}{2\sqrt{2}}$ a normalizing factor that is needed to convert a 3D positional error to a distance error \footnote{The derivation of $C$ is given in the appendix.}. The impact of different noise amplitudes on the final dose distribution is given in Section \ref{sec:robustness_noise}.

\section{Results}
\label{sec:results}
\subsection{Treatment plan optimization}

\begin{table}
    \centering
    \begin{tabular}{m{0.06\textwidth}|m{0.06\textwidth}m{0.06\textwidth}m{0.06\textwidth}m{0.06\textwidth}m{0.06\textwidth}||m{0.06\textwidth}m{0.06\textwidth}m{0.06\textwidth}m{0.06\textwidth}m{0.06\textwidth}}
        Pt. nb. & \multicolumn{5}{c||}{4D robust} & \multicolumn{5}{c}{Library of plans} \\\hline
		& \# spots & $D_{95}$ [\%] & $D_5$ [\%] & Homog-eneity  [\%] & Liver-CTV $D_{mean}$ [Gy] & \# spots & $D_{95}$ [\%] & $D_5$ [\%] & Homog-eneity [\%] & Liver-CTV $D_{mean}$ [Gy] \\\hline
        1 & 3936 &  97.3 & 104.8 & 92.5 & 3.9 & 11271 & 97.6 & 102.8 & 94.9 & 3.0\\
        2 & 2140 & 98.0 & 103.3 & 94.7 & 7.5 & 6292 & 99.0 & 102.4 & 96.6 & 6.5\\
        3 &  2178&  98.5 & 103.7 & 94.9 & 2.1 &  6573& 100.1& 103.7 & 96.5 & 1.6\\
        4 &  7615&  97.2 & 105.0 & 92.2 & 6.6 &  21226&  97.5 & 103.0 & 94.5 & 5.4\\
        5 &  4574& 98.2 & 104.2 & 93.9 & 3.2 & 12009& 99.8 & 101.6 & 98.2 & 3.0
    \end{tabular}
    \caption{Optimized treatment plans characteristics. Comparison between the 4D robust approach and the library of plans static dose characteristics and the number of spots.}
    \label{tab:treatment_plans}
\end{table}

\noindent For each of the five patients detailed in Table \ref{tab:patient}, we compute one 4D robust treatment plan and ten 3D robust treatment plans on each phase of the 4DCT. We report in Table \ref{tab:treatment_plans} the total number of spots in the 4D robust plan and the library of treatment plans, as well as key dose-volume histogram (DVH) metrics on the static doses.\\
\\
From those results, we can already conclude that the theoretically best achievable multi-gating solution surpasses the conventional 4D robust plan in both the target homogeneity and the surrounding OAR dose reduction. However, the total number of spots is, on average, roughly increased by a factor of 3, which remains acceptable considering that there are ten plans in the library. In the following sections, we simulate our approach dynamically on a continuous sequence of CTs to evaluate its capabilities.

\subsection{Choosing the distance metric}
\label{sec:distance_measure}
\noindent We compare two metrics: the Euclidean distance and DICE similarity coefficient in Table \ref{tab:compare_dist_dice} under perfect knowledge of the tumor position. First of all, we observe that our approach has a better dose homogeneity in the target as well as a lower mean liver-CTV dose in all cases compared to the 4D robust plan. However, treatment time is greater, as expected, for two reasons. First, there are more spots to deliver in our approach since there are multiple plans. Secondly, the duty cycle of the beam will always be lower than the one of the 4D robust approach, which is 100\% since the plan is delivered without interruption.\\
\\
Concerning the comparison between the use of the Euclidean distance between centers of mass and the use of the DICE similarity measure, the two approaches give qualitatively similar results. It is hard to give a definitive answer as the decision parameter, i.e. the Euclidean distance threshold and the DICE threshold, are not directly comparable. However, no approach seems to be clearly superior to the other. Hence, because the distance to the center of mass is a simpler metric, more easily clinically implemented, and has better threshold interpretability, we choose to keep this distance metric throughout the rest of the paper.

\begin{table}
\begin{tabular}{c|c|c|c|c|c}
\textbf{}    & \textbf{CTV $D_{95}$} & \textbf{CTV $D_5$} & \textbf{\begin{tabular}[c]{@{}c@{}}Homogen-\\ eity [\%]\end{tabular}} & \textbf{\begin{tabular}[c]{@{}c@{}}Mean Liver-\\ CTV dose\end{tabular}} & \textbf{\begin{tabular}[c]{@{}c@{}}Treatment\\ time [min]\end{tabular}} \\\hline
Distance 1mm & 59.0             & 61.89           & 95.18                                                            & 2.52                                                                    & 15.08                                                             \\
Distance 2mm & 58.88            & 61.60           & 95.46                                                            & 2.55                                                                    & 4.51                                                              \\
Distance 3mm & 58.41            & 61.89           & 94.20                                                            & 2.59                                                                    & 3.83                                                              \\\hline
DICE 0.75    & 57.98            & 62.95           & 91.72                                                            & 2.57                                                                    & 3.51                                                              \\
DICE 0.85    & 58.53            & 61.62           & 94.85                                                            & 2.64                                                                    & 3.90                                                              \\
DICE 0.95    & 58.97            & 61.53           & 95.73                                                            & 3.49                                                                    & 21.37                                                             \\\hline
4D robust    & 56.76            & 64.21           & 87.58                                                            & 3.33                                                                    & 2.19                                                             
\end{tabular}
\caption{Comparison between distance to the center of mass and DICE similarity measure for decision-making. Units are in Gy unless otherwise stated. The last row corresponds to the standard 4D robust model for completeness.}
\label{tab:compare_dist_dice}
\end{table}

\subsection{What image acquisition frequency do we need?}
\label{sec:image_frequency}
\begin{figure}
\centering
\begin{subfigure}{.5\textwidth}
  \centering
  \includegraphics[width=\linewidth]{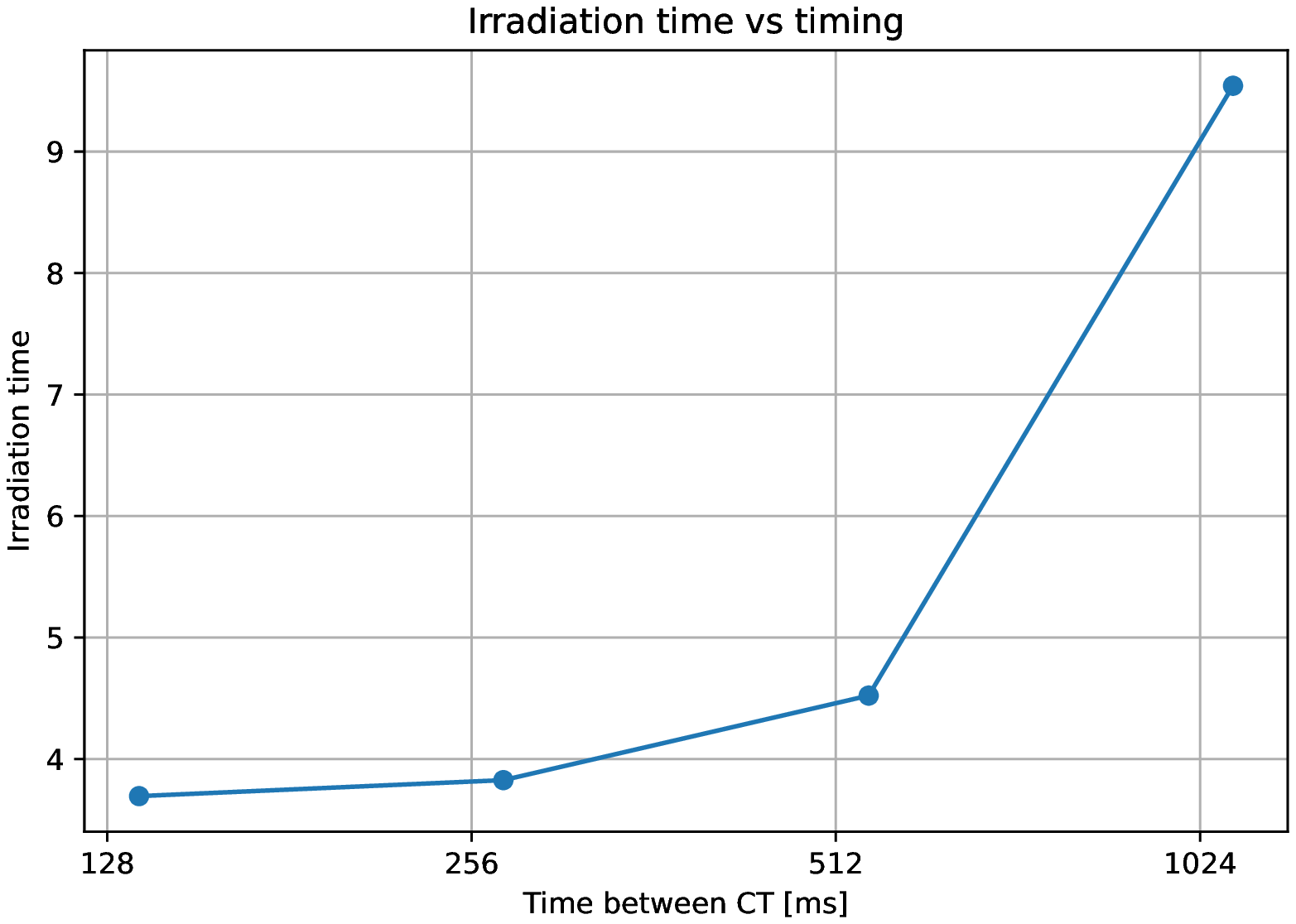}
  \caption{Treatment time analysis}
  \label{fig:sub1}
\end{subfigure}%
\begin{subfigure}{.5\textwidth}
  \centering
  \includegraphics[width=\linewidth]{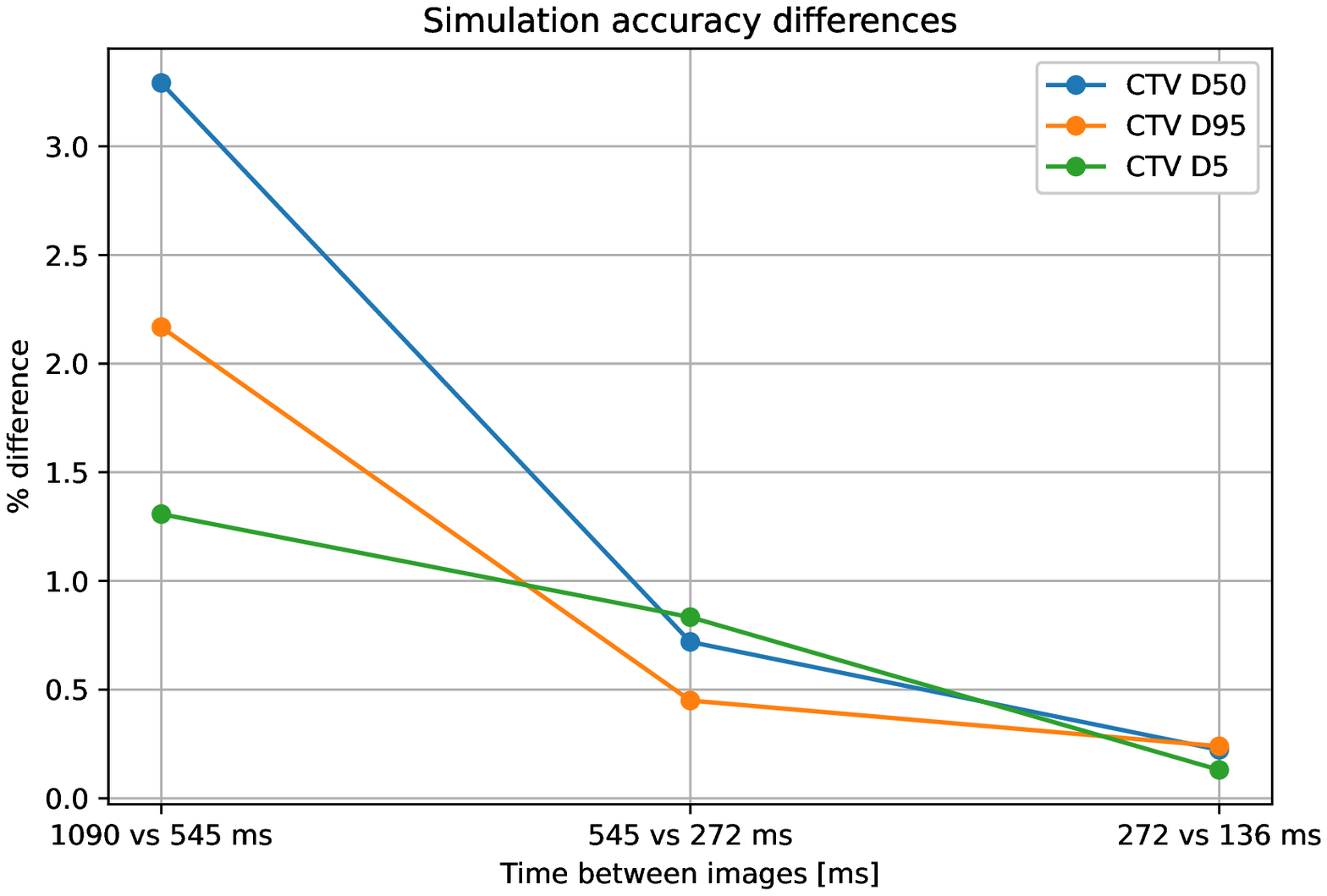}
  \caption{Simulation accuracy analysis}
  \label{fig:sub2}
\end{subfigure}
\caption{(a) Treatment timing as a function of the image rate. (b) Percentage difference between the values of three DVH metrics for different image rates.}
\label{fig:difference_timings}
\end{figure}

\noindent To answer this question, we compute three DVH metrics: $D_{95}$, $D_{50}$, and $D_5$ at different image acquisition frequencies. The period of acquisition was, in our case, 545 ms between images. We downsampled to 1090 ms and upsampled to 222.5 and 111.25 ms to study the impact of the acquisition rate.\\
\\
The results for the different image acquisition rates are depicted in Figure \ref{fig:difference_timings}.
In the right figure, we observe a difference of less than 1\% for the three DVH metrics between a period of 545 ms and 272ms. Therefore, 545 ms is a good enough image period for the simulation accuracy. However, in the left figure, we observe that an image rate of 545 ms leads to a longer treatment time than a 222.5 ms rate. This is because more decisions are taken, effectively reducing the duty cycle of the beam. Further increasing the imaging rate only decreases the treatment time by a tiny margin, most likely because the tumor movement becomes very small, which does not induce a change in the plans selected. Therefore, a period of around 225ms seems to be a good trade-off and will be kept as the default value for the rest of the simulations.


\subsection{Robustness to noise: analysis of one patient}
\label{sec:robustness_noise}
In this section, we study the impact of imperfect information on the tumor's position  as well as different distance thresholds for decision-making on the final dose. We compare five distance thresholds (1, 2, 3, 4, 5mm), used to select which plans will be delivered, under four distance error scenarios (0, 1, 2, 3mm) on patient 5 and report the results in Figure \ref{fig:p19_results} (the results for the other patients are available in the supplementary materials: Figures S5-8). For each threshold-error pair, we ran the simulations from five different starting images within the continuous sequence, giving us a statistical approximation of the uncertainty represented in the box plots. Three metrics are compared: $D_{95}$ and $D_5$ given in percentage of the prescription and the treatment time.\\

\begin{figure}[t!]
    \centering
    \includegraphics[scale=0.65]{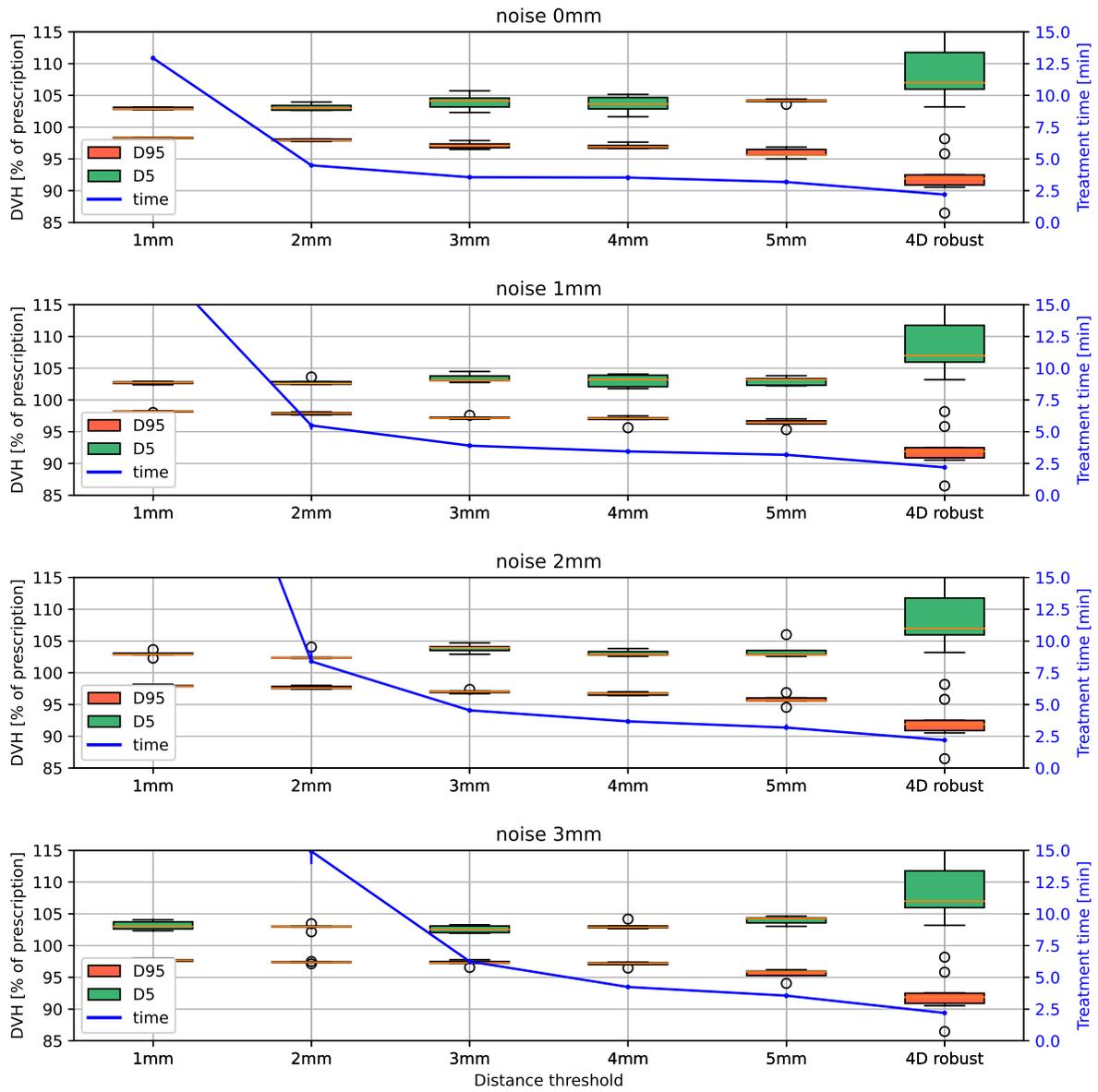}
    \caption{Patient 5 results (DVH characteristics and treatment time) for increasing noise amplitudes on the location of the tumor (top to bottom) and increasing distance thresholds (left to right) for selecting the plans in the library. The distance thresholds are only the initial thresholds, and they follow the increase scheme described in Section \ref{sec:delivery}, that is, the distance thresholds are increased to 5mm if no spot were irradiated in the last 10 seconds and further increased to 10mm if no spots were irradiated in the last 20 seconds. The 4D robust approach results are repeated on the far right of each subplot. They are not affected by the noise amplitudes since the plan is not synchronized with tumor motion. The boxplots and small vertical bars on the blue solid line account for uncertainties linked to the different starting times for the simulation on the CS.}
    \label{fig:p19_results}
\end{figure}

\begin{figure}
\centering
\begin{subfigure}{.5\textwidth}
  \centering
  \includegraphics[width=0.95\linewidth]{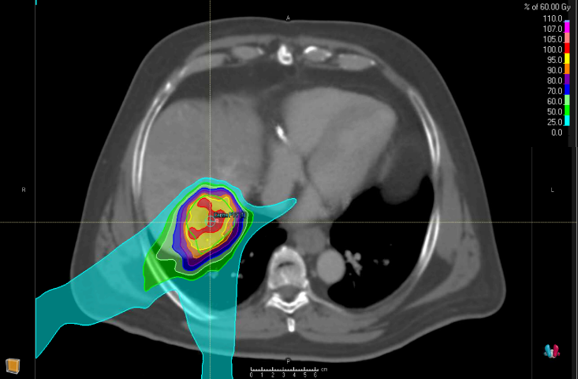}
  \caption{4D robust approach}
  \label{fig:dose_distrib_p5_4Drobust}
\end{subfigure}%
\begin{subfigure}{.5\textwidth}
  \centering
  \includegraphics[width=0.95\linewidth]{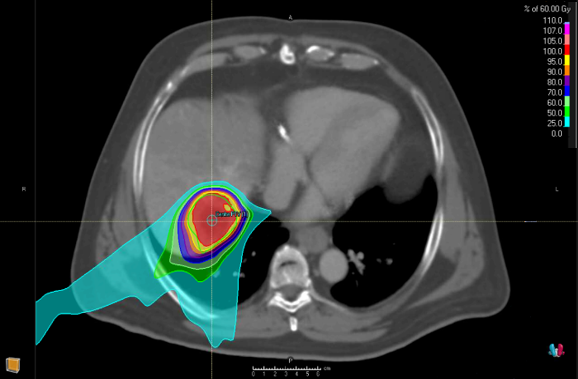}
  \caption{Library of plans approach}
  \label{fig:dose_distrib_p5_library}
\end{subfigure}
\caption{Comparison of spatial dose distributions for patient 5 under a noise amplitude of 2mm and distance threshold of 3mm.}
\label{fig:dose_distribution_patient5}
\end{figure}

\noindent Looking at the noise-free scenario, we observe that increasing the distance threshold decreases the treatment time and slowly decreases the homogeneity (represented by the gap between the $D_{95}$ and $D_5$ metric). We also observe that even at a distance of 5mm, we are still delivering a more homogeneous dose to the target compared to the 4D robust plan. Moreover, the uncertainty in the final dose due to the different starting times is low compared to the 4D robust plan.\\
\\
Looking at the impact of the noise amplitude on the outcome of the simulated treatment, we observe that the dose is slightly deteriorating but staying at a reasonable target coverage, still better than the 4D robust plan. However, the treatment time is impacted by the noise, especially for the 1 and 2mm distance thresholds. Hence, we should aim for a distance threshold of at least 3mm for these parameters. For the location error, depending on the type of tracking chosen, we can expect different levels of noise amplitudes. In the case of fiducial-based tumor tracking, we can expect a sub-millimeter accuracy on 2D images and an error within 1mm for three-dimensional calculation error, as reported in a recent study on fiducial markers \cite{miyamoto2019quantitative}, hence a noise amplitude $<$ 2mm. For markerless tracking, a recent paper achieved a tracking accuracy of $1.64 \pm 0.73 mm$ \cite{hirai2019real}; hence, a noise amplitude of 2mm seems to be a correct expected error for today's tumor tracking accuracy. For the patient analyzed in Figure \ref{fig:p19_results}, this would lead to a treatment time below 5 minutes, which still seems reasonable. The spatial distribution of the dose for this patient under a noise amplitude of 2mm is compared to the 4D robust approach in Figure \ref{fig:dose_distribution_patient5}.

\subsection{Comparison between all patients}
\label{sec:results_all_patients}
\noindent In this section, we compare the results of the simulation study on the five patients described in Section \ref{sec:patient_data} with a distance threshold of 3 mm and a noise amplitude of 2 mm. During the initialization phase, the number of plans selected, satisfying the constraint $O_{p_i}>5\%$ (see equation \ref{eq:Opi}), are respectively 4, 10, 7, 5, and 10 for patients 1 to 5. Results of the treatment delivery are depicted in Figure \ref{fig:all_patients} (Figures S9-13 in the supplementary materials provide additional details on the cumulative DVH results). We observe that the dose is more homogeneous in the target for all patients with our approach except for patient 2, for which the homogeneity is similar (4D robust plan is 1\% better). On average, the homogeneity increase was 5\% for the library of treatment plans approach. This can be quite surprising because the 4D robust plan has, by design, greater margins around the target and should therefore lead to a more conformal dose. However, this is not necessarily the case when motions not present in the planning 4DCT are encountered during treatment delivery, which we noticed to be the case for a non-negligible part of the treatment delivery on all patients in this case study.\\


\begin{figure}[t!]
\centering
\begin{subfigure}{\textwidth}
  \centering
  \includegraphics[width=\linewidth]{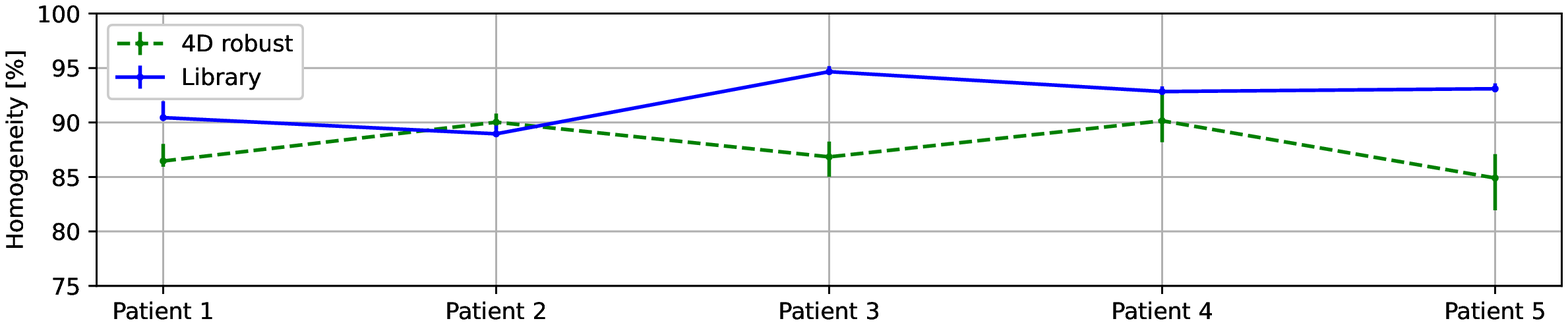}
  \caption{Homogeneity in the target volume (computed according to equation \eqref{eq:homogeneity}). The vertical bars describe the interquartile range (IQR), and the continuous line represents the median value. The closer we are to 100\%, the better.}
  \label{fig:all_homogeneity}
\end{subfigure}
\begin{subfigure}{\textwidth}
  \centering
  \includegraphics[width=\linewidth]{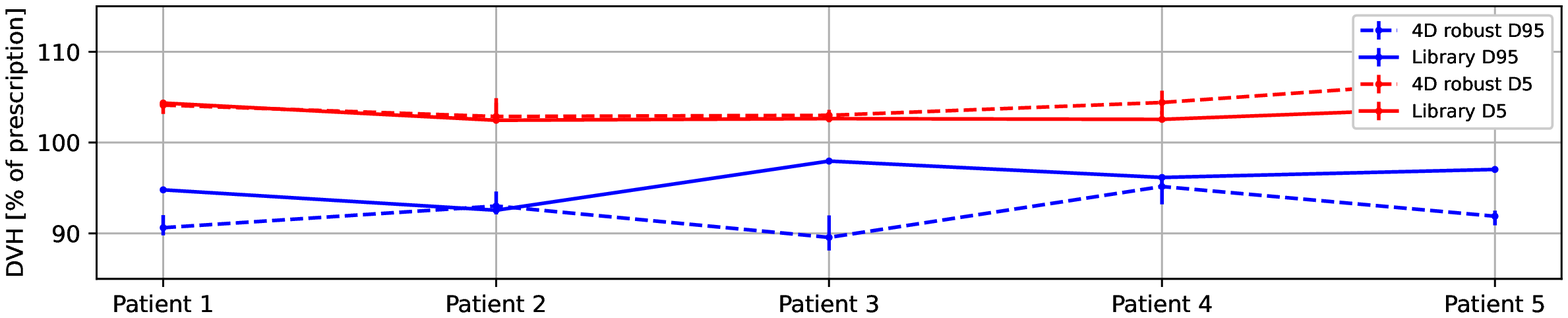}
  \caption{$D_{95}$ and $D_{5}$ metrics in the target volume in percentage of the prescription.}
  \label{fig:all_DVH}
\end{subfigure}
\begin{subfigure}{\textwidth}
  \centering
  \includegraphics[width=\linewidth]{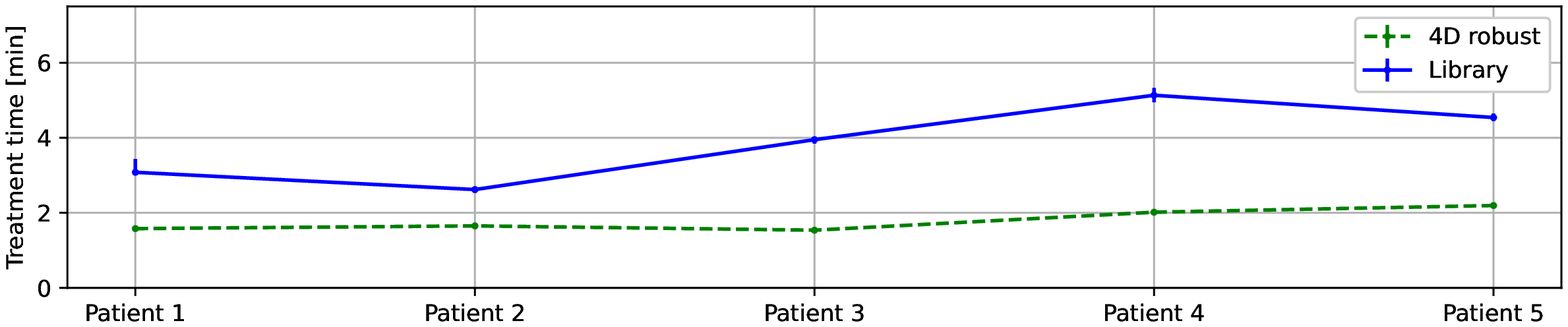}
  \caption{Time required for delivering the treatment.}
  \label{fig:all_time}
\end{subfigure}
\begin{subfigure}{\textwidth}
  \centering
  \includegraphics[width=\linewidth]{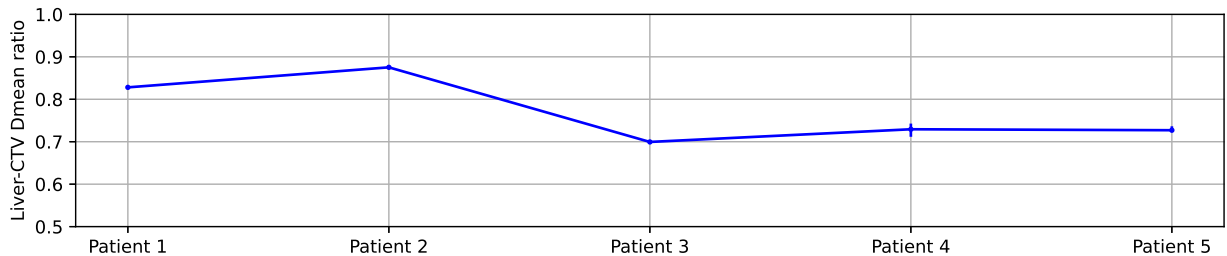}
  \caption{Ratio between the mean doses delivered in the Liver-CTV volume (CTV subtracted from the liver) of the library of plans and the 4D robust approach. A value below 1 indicates an improvement compared to the 4D robust approach.}
  \label{fig:all_liver_dmean}
\end{subfigure}
\caption{Simulation results of the treatment delivery on five patients for the 4D robust approach and the library of plans approach. A noise amplitude on the distance to the tumor of 2mm is applied, and a distance threshold of 3mm is chosen. The vertical bars indicate uncertainty linked to the different starting times for the simulation on the CS.}
\label{fig:all_patients}
\end{figure}

\noindent Since margins are smaller with our approach, we also expect that the dose to the organs at risk will be reduced, which is confirmed in the last plot in Figure \ref{fig:all_patients}. We computed the ratio between the mean Liver-CTV dose of our approach and the mean Liver-CTV dose of the 4D robust model. We observe that for all patients, the mean dose of uninvolved liver is reduced by 23\% on average and up to 30\%.\\
\\
Finally, the treatment time is inevitably increased in our approach. On average, it is increased by roughly a factor of 2, which lead to a mean treatment time of 3.8 minutes and a maximum treatment time of 5.1 minutes, which still seems reasonable. In this treatment time, we also included the 30-second initialization phase described in Section \ref{sec:delivery}.

\section{Discussion}
\paragraph{Adapting treatment in real-time} In this study, we have seen that real-time adaptation of the treatment based on the current anatomy using a library of treatment plans helps to achieve a better dose homogeneity in the target and a significant decrease in the surrounding OARs compared to a state-of-the-art 4D robustly optimized treatment plan. This comes at the expense of an increased treatment time. On average, we manage to limit this increase to a factor of 2. Considering that we have ten treatment plans and an active beam delivery, i.e. a duty cycle strictly lower than 1, this is considered a quite good performance. This was possible because of sparser treatment plans enforced by equation \eqref{eq:constraint}. Nevertheless, the tumors in this case study were relatively small, and big tumors might pose a problem in terms of increased treatment time. Improvements on treatment plan sparseness are still possible, hence further decreasing the treatment time. One of the possibilities would be regularizing the objective function \eqref{eq:obj} by adding a penalty term on the weights via a $\ell_1$ norm. We leave this for a future study. Another approach that could increase the duty cycle is a relative weighting of the plans according to the time spent in each phase. Indeed, it is possible that one phase is available for a longer amount of time than another phase, which could be evaluated during the 30-second initialization phase. This could also provide a gain in dose conformity because our approach prioritizes the less-delivered plan over the most similar one.

\paragraph{Tumor tracking} Although we did not restrict ourselves in this study to a specific imaging modality, tracking tumors seems to be best handled with fluoroscopy in today's technology, mainly because this technology is already mostly integrated into today's proton therapy systems. Indeed, nowadays, proton therapy machines are usually either equipped with a Cone-beam CT imaging system or 2D orthogonal radiographs and can be used for real-time image guidance \cite{kruse2018proton, korreman2015image,vemprala2018real}. Moreover, the results obtained in Section \ref{sec:results_all_patients} assume a distance error of 2 mm, which is what today's tumor tracking technology can already achieve \cite{miyamoto2019quantitative, hirai2019real}. Should our approach be implemented in a proton therapy system, no hardware changes would be needed for the imaging system. However, some changes would be necessary for the treatment delivery system, which would need to be synchronized with the anatomical motion and selection of treatment plans. Such a modular dose delivery system prototype has already been successfully implemented for scanned ion beam \cite{lis2020modular} and characterized at GSI Helmholtzzentrum für Schwerionenforschung GmbH and the Centro Nazionale di Adroterapia Oncologica (CNAO).

\paragraph{System latency and tumor motion prediction} We have empirically demonstrated that an image acquisition with a period of roughly 250ms was enough for making decisions on the active delivery of the treatment. Decreasing this period only marginally decreases treatment time below this value. However, we neglected the latency of the system, defined as the duration between the start of the image acquisition and the beam delivery or beam off. The system latency includes the image acquisition duration, time for processing the image, time for selecting the treatment plan based on the distance to the current tumor position, and time for switching the beam ON or off. If this time lag is too large, this could affect the accuracy of the dose delivery by irradiating while the target has already moved out of the beam path. AAPM Task Group 264 for the safe clinical implementation of MLC tracking in radiotherapy recommends as a minimum requirement an average system latency $\leq$ 500ms \cite{keall2021aapm}. This is well over system latencies currently obtained for gated photon and proton therapy treatments. In proton therapy using PBS with real-time imaging and amplitude-based gating, Hokkaido University Hospital PT center reported a system latency of 66ms \cite{shimizu2014proton}. Another group recently commissioned a fluoroscopic-based real-time markerless tumor tracking system for carbon-ion PBS and reported an overall system latency of 70-110ms \cite{mori2019commissioning}. For surrogate-based tracking with optical and electromagnetic technologies, the worst reported system latency was 31 ms \cite{fattori2017monitoring}. In our approach, we could expect a similar latency as PBS with gating based on fluoroscopy imaging, i.e. 100ms up to 200ms if the selection of treatment plans takes more time. Because tumor motion is predictable \cite{verma2010survey}, the system latency can be compensated with a prediction filter predicting future motion. A larger number of predictive models have been developed in the last decade \cite{verma2010survey, ernst2013evaluating, johl2020performance, lin2019super}. In the latest study \cite{johl2020performance} comparing several predictive models, the authors reported a normalized RMSE\footnote{The normalized RMSE is computed as $\frac{RMSE(y_{orig},y_{pred})}{RMSE(y_{orig},y_{delayed})}$} $<$ 0.05 for a prediction horizon of 160 and 480 ms with a linear filter, meaning that prediction is 95\% better than without a prediction (1 corresponds to no improvement and 0 to a perfect prediction). This corresponds to a sub-mm accuracy even for high motion amplitudes. However, the prediction filters in the study used a sampling frequency of 25Hz, while we used a sampling frequency $\sim$ 4Hz. A hybrid motion tracking using fluoroscopy and an external surrogate might help achieve a higher sampling rate and increase the accuracy of motion prediction.

\paragraph{Distance metrics} We found in Section \ref{sec:distance_measure}, that the center of mass of the tumor was a good proxy to decide which plans to select when acquiring an image. We also considered the DICE similarity index, which provides richer information on the tumor's location but did not significantly improve the performance. However, those two metrics only look at tumor information, not the anatomy globally. Other measures that take the global anatomy into account could be tested, such as the mutual information between the current image and the images of the 4DCT. We leave this as future work.

\paragraph{Application to lung cancers} In this simulation study, we only looked at liver cancers simply because we had real breathing MRI acquisitions for those patients at our disposal. However, we expect our approach to display even better performance in lung cancer patients because the effect of motion is amplified due to the lower tissue densities encountered for those cancers, making a single treatment plan more prone to dose heterogeneities than our library of treatment plans approach.

\paragraph{Choosing the number of breathing phases:} The number of breathing phases and treatment plans in our approach was simply determined by the number of phases in the planning 4DCT. However, we did not investigate the outcome of a higher or lower number of phases. Taking more phases would probably not be a good idea in terms of the negative impact on the treatment time (because more plans would need to be delivered), but fewer phases might offer a better trade-off between treatment time and dose conformity.

\paragraph{Comparison with gated delivery on maximum exhalation or inhalation} Our approach is essentially an extension of amplitude-based gating for multiple plans. If the breathing motion during treatment delivery is similar to the one of the 4DCT, then gated delivery on maximum exhalation or inhalation and our approach should both lead to approximately the same treatment time (for the same choice of parameters). However, our approach should lead to a better homogeneity in the target volume because of the intrinsic volumetric rescanning induced by delivering multiple plans. To obtain the same dosimetric outcome, a conventional gated delivery would need several rescanning, which would increase the treatment time considerably. Another disadvantage of gating is the possibility that the plan can never be delivered if there is a baseline shift between planning and delivery. This is the case for patient 1 (see Figure S-1 in the supplementary materials) in our case study. A treatment plan optimized on max inhalation would be unable to be delivered in that case, while our method allows adapting the plan to be delivered during the initialization phase. Finally, for a gated plan where delivery window is set across several motion phases with the target defined as an ITV, the difference with our approach would be larger margins for the gating solution and therefore higher dose for surrounding organs.

\section{Conclusion}
\noindent We developed a real-time image-guided approach for treating moving tumors with a library of treatment plans optimized on each phase of a planning 4DCT. Our approach, simulated with the current accuracy of tumor tracking technology, allows to reduce the dose in the surrounding OARs and improve the dose homogeneity in the target compared to a state-of-the-art 4D robust treatment plan. This comes at the expense of an increased  treatment time but remains below an acceptable level of 4 minutes on average.

\newpage
\section*{Appendix}
\addcontentsline{toc}{section}{\numberline{}Appendix}
\section*{Multivariate Gaussian noise and distance error}
\noindent To add a Gaussian noise on a $N$-dimensional position that represents an average distance error, we need to look at the expected value of the norm of the noise vector.\\
\\
Let us define $\mathbf{p} \in \mathbb{R}^N$, a position in $N$ dimensions. If we add a Gaussian noise $\mathbf{e}\sim \mathcal{N}(\mathbf{0},\boldsymbol{\Sigma})$ with $\boldsymbol{\Sigma}=\sigma^2 \mathbf{I}$ to $\mathbf{p}$, we obtain a noisy position $$\mathbf{y}=\mathbf{p} + \mathbf{e}$$
How can we choose $\sigma$ so that the average distance error is equal to $d$? We need to look at the expected value of the norm of the noise vector $\mathbf{e}$. Mathematically, that is
\begin{align}
\mathbb{E}_{\mathbf{e}\sim \mathcal{N}(\mathbf{0},\sigma^2\mathbf{I})}\left\lbrace \| \mathbf{e} \|_2 \right\rbrace &= \mathbb{E}_{\bar{\mathbf{e}}\sim \mathcal{N}(\mathbf{0},\mathbf{I})}\left\lbrace \| \sigma \bar{\mathbf{e}} \|_2 \right\rbrace \\
&= \sigma  \mathbb{E}_{\bar{\mathbf{e}}\sim \mathcal{N}(\mathbf{0},\mathbf{I})}\left\lbrace \| \bar{\mathbf{e}} \|_2 \right\rbrace \\
&= \sigma  \mathbb{E}_{\bar{\mathbf{e}}\sim \mathcal{N}(\mathbf{0},\mathbf{I})}\left\lbrace \sqrt{\sum_{i=1}^N e_i^2} \right\rbrace
\end{align}
We thus have the square root of a sum of squares of $N$ standard normal random variables, which is a Chi-distribution for which the mean is given by \cite{evans2011statistical}
\begin{equation}
    \mu=\sqrt{2}\frac{\Gamma\left(\frac{N+1}{2}\right)}{\Gamma\left(\frac{N}{2}\right)}
\end{equation}
where $\Gamma$ is the gamma function and $N$ the dimension. For $N=3$, we have $\mu=\frac{\sqrt{2}\Gamma(2)}{\Gamma(3/2)}=\frac{2\sqrt{2}}{\sqrt{\pi}}$. Hence, for a 3-dimensional position, if we want the average distance error to be equal to $d$, we must set $\sigma=\frac{\sqrt{\pi}}{2\sqrt{2}}d$.

\clearpage
\section*{References}
\addcontentsline{toc}{section}{\numberline{}References}
\vspace*{-20mm}





\bibliography{./bibliography}      



\bibliographystyle{./medphy.bst}    


\end{document}